\documentclass[11pt,a4paper]{article}
\usepackage{graphicx}

\title{\bf Clinical trials with rescue medication applied according to a deterministic rule}
\author{Gerd K Rosenkranz\thanks{Address: Gerd K Rosenkranz, Center for Medical Statistics, Informatics and Intelligent Systems, Medical University of Vienna, Spitalgasse 23, A-1090 Vienna, Austria. Email: {\tt gerd.rosenkranz@meduniwien.ac.at}}
\\Medical University of Vienna, Vienna, Austria}
\date{Version of \today}
\begin{document}
\maketitle
\begin{abstract}
Clinical trials in specific indications require the administration of rescue medication in case a patient does not sufficiently respond to investigational treatment. The application of additional treatment on an as needed basis causes problems to the analysis and interpretation of the results of these studies since the effect of the investigational treatment can be confounded by the additional medication. Following-up all patients until study end and capturing all data is not fully addressing the issue.
We present an analysis that takes care of the fact that rescue is a study outcome and not a covariate when rescue medication is administered according to a deterministic rule. This approach allows to clearly define a biological effect. For normally distributed longitudinal data a practically unbiased estimator of the biological effect can be obtained. The results are compared to an ITT analysis and an analysis on all patients not receiving rescue.
\end{abstract}
\section{Introduction}
In clinical trials in specific indications it is necessary to provide patients  the opportunity to receive an established medication, either alone or in combination with study treatment, in case of insufficient response. This additional treatment is called rescue medication, or sometimes reliever medication. In diabetes trials patients are to receive rescue medication in case blood glucose does not decrease over time (see for example \cite{Bailey_etal2013}). In trials in chronic obstructive pulmonary disease (COPD)~\cite{Buhl_etal2011, Korn_etal2011} rescue may be given to prevent exacerbations.

The rules according to which the need for rescue medication is determined varies between indications. For diabetes trials, these rules are often precisely defined. For example study~\cite{Bailey_etal2013} required open-label piaglitazone or acarbose for fasting plasma glucose concentrations above 15 mmol/L (week 4--8), 13.3 mmol/L (week 8--12) or 11.1 mmol/L (week 12--24). For rescued patients, measurements obtained after initiation of rescue medication were excluded from the efficacy analysis which consisted of an analysis of covariance with last observation carried forward for missing data.

In COPD trials, the patient determines the need for rescue medication and applies it himself. Since short acting medication is used, the impact on the results of the study outcome is considered minimal as long as rescue is not taken on the day of an assessment. The primary analysis of the main endpoint, change of forced expiratory volume within one second (FEV1), is an ITT-analysis including all patients  irrespectively of rescue medication. The frequency and amount of the latter is usually analyzed as a secondary endpoint to support the results of the primary analysis~\cite{Buhl_etal2011, Korn_etal2011}.

In general, the application of additional treatment not delivered according to random assignment but for a treatment emergent reason causes problems for analysis and interpretation of the results. Allowing or requesting rescue medication confounds the effect of the treatment under investigation. The administration of rescue medication is strictly speaking a study outcome, not a covariate, which in turn may have an impact on the observed outcome.

Recent literature encourages to follow-up all randomized subjects even if they alter the randomly assigned treatment~\cite{NAS2010}. This is not a principal problem for patients receiving rescue. However, it is not a priori clear whether and how to use the data obtained after rescue in a meaningful way in the statistical analysis of the trial (see~\cite{ONeill_Temple2012, White_etal2012}).

One approach to analyze data from trials calling for rescue is a strict ITT analysis, i.e., an analysis that considers only the assigned treatment and ignores the administration of rescue. This can be sensible approach when rescue is needed by the majority of patients at some point in time during the trial but is applied only for a short amount of time without interruption of the treatment under investiagtion. By following this principle one preserves the benefit of randomization in creating groups that do not differ systematically on any factors except those assigned in an trial. However, such an analysis would estimate the effect of assigning control or treatment, but not necessarily the biological effect of the latter over the former. If the biological effect of the investigational treatment alone is of interest, the ITT analysis is not appropriate~\cite{Sheiner_Rubin1995, Rosenkranz2013}.

One could also consider data after rescue as missing and impute them by carrying the last observation forward (as in \cite{Bailey_etal2013}) or by a more appropriate procedure (see~\cite{NAS2010}). However, it is not obvious which effect would be estimated by such an analysis.

The notion of a per protocol analysis is not straightforward in the context of rescue medication. When rescue is administered according to protocol, receiving rescue would not constitute a protocol deviation and the respective patients would still be part of the per-protocol set. Alternatively one could compare only data from patients that did not receive rescue medication. However, such an analysis would consider observations that have been selected according to their response to different treatments. Since this analysis would not compare like with like, the biological effect can also not be estimated appropriately in this way~\cite{Sheiner_Rubin1995, Rosenkranz2013}. For a quantitative account of the impact of rescue on standard analyses see White et. al~\cite{White_etal2001}.

Eventually one could also interpret administration of rescue medication as treatment failure and compare the number of patients that need rescue medication (or likewise the time to rescue medication) or the amount of rescue taken. However such an analysis does not provide an estimate of an effect on the primary outcomes of the study and may therefore be best used as a supportive analysis (like in COPD trials).

A different way to think about how to obtain an estimate of the biological effect is by considering potential outcomes~\cite{Neyman1990} and principal strata~\cite{Frangakis_Rubin2002}. These notions have been introduced to investigate the impact of non-compliance to assigned treatment. The key idea is to stratify on an intermediate outcome, in our case the administration of rescue. However, patients are not stratified according to the \textit{observed} administration of rescue, which is an outcome affected by the treatment assigned at baseline. Rather they are classified according to whether they would (or would not) need rescue under investigational treatment or control.

In the context of rescue medication, one can consider four strata of patients: those who do not need rescue regardless of assigned treatment (abbreviated 00), those who need rescue when assigned control but not when assigned treatment (10), those who need rescue when assigned treatment but not when assigned control (01) and those in need of rescue no matter how treated (11). If control treatment is a placebo, one may argue that stratum (01) is empty since patients not responding to investigational treatment are not responding to placebo either since the former is a placebo as well. This approach has been applied under a likelihood~\cite{Shaffer_Chinchili2003} as well as a Bayesian paradigm~\cite{Shaffer_Chinchili2004} to obtain estimates for the biological effect in subjects that do not fail under either treatment option (stratum 00).  However, univariate methods were considered that would necessitate to define a summary outcome for studies with repeated measures.

We consider a simple longitudinal setting consisting of 2 visits only and an explicit (deterministic) rule for the administration of rescue. The rule basically states that if the values at an earlier visit are lower or higher than acceptable, the subject should receive rescue medication. The endpoint to assess an effect of the investigational treatment will be captured at the final (second) visit of the trial. The only stratum that provides an non-confounded estimator of the treatment effect is patients that do not need rescue under treatment nor control (00). Under plausible distributional assumptions, the biological effect of treatment over control in the 00 stratum can be estimated without having to use computationally intensive methods like EM algorithms or Markov Chain Monte Carlo techniques.

After introducing notation and concept of potential outcomes and their relationship to observations, a practically unbiased estimator of the biological effect of interest is presented. Its small sample properties are investigated by simulation. Some questions for further investigations are discussed.
\section{Observations and potential outcomes}
We consider a parallel group trial where experimental treatment $(Z=1)$ is to be compared with control $(Z=0)$ treatment. The treatment is to be randomly assigned to $1\le i\le n_j, (j=0,1)$ patients who provide observations $(Y_{i1},Y_{i2})$ from two visits. For the discussion we assume that high values are beneficial and low values indicate lack of efficacy. The rule for applying rescue is that if the observed values are too small at the first visit after treatment initiation, i.e., $Y_{i1}\le c$ for some predefined critical value $c$, the subject has to receive rescue medication, denoted by $R_i=1$. $R_i=0$ means that no rescue is necessary. As a consequence, the outcome at the second visit will be affected not only by the treatment assigned at baseline, but also by the rescue medication administered after the first visit. Not to complicate matters further, we assume complete data at each visit, i.e., $(Y_{i1},Y_{i2})$ are observed for all patients and investigators and patients adhere to the rules for administration of rescue medication.

In the following we use the notion of potential outcomes. We drop the index accounting for the subject number whenever possible to ease notation. Let $Y_1(z)$ be the potential outcome of a subject at visit~1 had treatment $z$ been assigned at baseline. Let $R(z)$ denote the potential rescue medication of a subject assigned to treatment $z$. Likewise, let $Y_2(z,r)$ denote the potential outcome of a subject at visit~2 assigned treatment $z$ at baseline and having received rescue medication $r$ at visit~1. The biological effect of treatment $z$ on the outcomes at visit~1 is given by $E_1=E[Y_1(1)]-E[Y_1(0)]$, and the biological effect of treatment $z$ at visit~2 in stratum $(i,j)$ is $E_2(i,j)=E[Y_2(1,i)]-E[Y_2(0,j)]$. Specifically in the stratum of subjects who do not need rescue under either treatment it is $E_2(0,0)=E[Y_2(1,0)]-E[Y_2(0,0)]$. An artificial example is displayed in Table~\ref{t:ps}.
\begin{table}
\begin{center}
\begin{tabular}{|c|c|c|c|c|c|c|}
\hline
  \% & Principal & \multicolumn{2}{c|}{Control} & \multicolumn{2}{c|}{Treatment} & Effect \\
  \cline{3-6}
     & stratum   & $R(0)$ & $E[Y(0,R(0))]$ & $R(1)$ & $E[Y_2(1,R(1))]$ &  \\
  \hline
  60 & 00 & 0 & 0 & 0 & 1 & 1 \\
  10  & 01 & 0 & 0 & 1 & 3 & 3 \\
  10 & 10 & 1 & 1 & 0 & 1 & 0\\
  20 & 11 & 1 & 1 & 1 & 3 & 2\\
  \hline
\end{tabular}
\end{center}
\caption{Example of expected outcomes in principal strata for a trial requiring rescue medication}\label{t:ps}
\end{table}
\par The left column shows the percentage of patients in the sample belonging to the respective principal strata. The causal effects for the respective strata are provided in the rightmost column.  Of course, in reality, these proportions and the respective strata means are not observable. There is a benefit for patients in all but one principal stratum from receiving treatment making treatment beneficial for 80\% of subjects. The ITT effect is $0.6\times 1+0.1\times 3+0.2\times 2=1.3$ while the effect in patients not requiring rescue is~1.

To obtain estimates of the biological effect $E_2(0,0)$, a connection between the potential outcomes and the observations $Y_1, Y_2, R$ has to be established. Obviously,
\begin{eqnarray*}
Y_1 &=& ZY_1(1)+(1-Z)Y_1(0)\\
R &=& ZR(1)+(1-Z)R(0)\\
Y_2 &=& ZY_2(1,R(1))+(1-Z)Y_2(0,R(0))\\
Y_1Y_2 &=& ZY_1(1)Y_2(1,R(1))+(1-Z)Y_1(0)Y_2(0,R(0))\\
\end{eqnarray*}
The last equation follows from $Z(1-Z)=0$, $Z^2=Z$ and $(1-Z)^2=1-Z$.
Under the assumption that $Z$ is assigned randomly to subjects, $Y_1(z)$ and $R(z)$ are independent of $Z$ and therefore
\begin{eqnarray*}
E[Y_1|Z=z] &=& E[Y_1(z)]\\
E[R|Z=z] &=& E[R(z)].
\end{eqnarray*}
This can be seen for $R$ from
$$E[R|Z=1]=E[ZR(1)+(1-Z)R(0)|Z=1]=E[R(1)|Z=1]=E[R(1)].$$
As a result, the conditional expectation of the observations $Y_1$ and $R$ given $Z=z$ equals the expectation of their respective counterfactual outcomes. This is no longer true for $Y_2$, since $Y_2$ is confounded by $R$. In fact
\begin{eqnarray*}
\lefteqn{E[Y_2|Z=1,R=0]}\\
&=& E[ZY_2(1,R(1))+(1-Z)Y_2(0,R(0))|Z=1,R=0]\\
&=& E[Y_2(1,R(1))|Z=1,R=0]\\
&=& E[Y_2(1,R(1))|R(1)=0]\\
&=& E[Y_2(1,0)|R(1)=0]\\
&=& E[Y_2(1,0)|Y_1(1)>c]
\end{eqnarray*}
Similarly one can show $E[Y_2|Z=0,R=0]=E[Y_2(0,0)|Y_1(1)>c]$, and therefore
\begin{equation}\label{e:exp2}
E[Y_2|Z=z,R=0]=E[Y_2(z,0)|Y_1(z)>c]
\end{equation}
confirming that an estimator of $E[Y_2|Z=z,R=0]$ is not appropriate as an estimator of $E[Y_2(z,0)]$.
\section{A biological effect estimator}
To obtain an estimator of $E[Y_2(z,0)]$ from $E[Y_2|Z=z,R=0]$, a correction is necessary. It is shown in this section that such a correction can be obtained for normally distributed variables. Assume that $(Y_1(z),Y_2(z,r))$ has a bivariate normal distribution with means
\begin{eqnarray*}
\mu_1(z)&=&E[Y_1(z)]=\alpha_1+\beta_1z\\
\mu_2(z,r)&=&E[Y_2(z,r)]=\alpha_2+\beta_2z+\gamma r+\delta zr
\end{eqnarray*}
and covariance matrix
$$\Sigma(z,r)=\left(\begin{array}{cc}\sigma_{11}^2(z)&\sigma_{12}(z,r)\\\sigma_{12}(z,r)&\sigma_{22}^2(z,r)\end{array}\right)$$
It should be noted that the decision on rescue at visit 1 can be based on an endpoint different from the one on which the efficacy assessment is based on at visit~2.

Under the model above, $\beta_2=\mu_2(1,0)-\mu_2(0,0)$ is the causal effect of $z$ at visit 2 in the principal stratum of subjects that do not require rescue under any assigned treatment. Let $\gamma_{12}(z,r)=\sigma_{12}(z,r)\sigma_{11}^{-2}(z)$ and recall that for bivariate normal variables
$$ E[Y_2(z,r)|Y_1(z)]=\mu_2(z,r)+\gamma_{12}(z,r)\{Y_1(z)-\mu_1(z)\} $$
holds. Let $\phi$, $\Phi$, and $\lambda$ denote the pdf, cdf and hazard function of a standard normal variable, respectively. With $\eta(z)=[c-\mu_1(z)]/\sigma_{11}(z)$, it follows that
\begin{eqnarray*}
E[Y_1(z)|Y_1(z)>c]&=&\mu_1(z)+\sigma_{11}(z)\phi(\eta(z))[1-\Phi(\eta(z))]^{-1}\\
&=&\mu_1(z)+\sigma_{11}(z)\lambda(\eta(z))
\end{eqnarray*}
Then
\begin{eqnarray}
\lefteqn{E[Y_2(z,0)|Y_1(z)>c]}\nonumber\\
&=&E[E[Y_2(z,0)|Y_1(z)]|Y_1(z)>c]\nonumber\\
&=&E[\mu_2(z,0)+\gamma_{12}(z,0)\{Y_1(z)-\mu_1(z)\}|Y_1(z)>c]\nonumber\\
&=&\mu_2(z,0)+\gamma_{12}(z,0)\{E[Y_1(z)|Y_1(z)>c]-\mu_1(z)\}\nonumber\\
&=&\mu_2(z,0)+\gamma_{12}(z,0)\sigma_{11}(z)\lambda(\eta(z))\label{e:cexp}
\end{eqnarray}
From equations (\ref{e:exp2}) and (\ref{e:cexp}) one obtains
\begin{equation}\label{e:mu2}
\mu_2(z,0)=E[Y_2|Z=z,R=0]-\gamma_{12}(z,0)\sigma_{11}(z)\lambda(\eta(z)).
\end{equation}
The conditional expectation of the observed value equals the expectation of the potential outcome if $\sigma_{12}(z,0)=0$, in which case observations from different visits are independent and the probability of rescue medication depends only on the randomly assigned treatment $z$ and is thus independent of the potential outcome. In this case, the causal effect $\beta_2$ equals  $E[Y_2|Z=1,R=0]-E[Y_2|Z=0,R=0]$. The latter is also the case if $\sigma_{12}(z,0)\lambda(\eta(z))$ does not depend on the assigned treatment $z$ which is a somewhat artificial condition.
\begin{table}[t]
\begin{center}
\begin{tabular}{cccccc|ccc}
\hline\hline
\multicolumn{6}{c|}{}&\multicolumn{3}{c}{Estimator of $\beta_2$}\\
$\alpha_1$ & $\beta_1$ & $\alpha_2$ & $\beta_2$ & $\gamma$ & $\delta$ & ITT & Conditional & Corrected\\
\hline
1 & 0 & 0 & 0 & 0 & 0 & 0.001 (0.201) & 0.001 (0.200) & 0.001 (0.200)\\
0 & 1 & 0 & 0 & 0 & 0 & -.003 (0.195) & -.226 (0.205) & -.003 (0.205)\\
0 & 0 & 1 & 0 & 0 & 0 & -.002 (0.205) & -.001 (0.221) & -.001 (0.221)\\
0 & 0 & 0 & 1 & 0 & 0 & 1.001 (0.201) & 1.000 (0.220) & 1.000 (0.220)\\
0 & 0 & 0 & 0 & 1 & 0 & 0.001 (0.179) & 0.001 (0.219) & 0.001 (0.219)\\
0 & 0 & 0 & 0 & 0 & 1 & 0.308 (0.191) & -.002 (0.219) & -.002 (0.219)\\
0 & 0 & 0 & 0 & 1 & 1 & 0.307 (0.192) & -.000 (0.219) & -.000 (0.219)\\
0 & 1 & 0 & 1 & 0 & 0 & 1.003 (0.199) & 0.781 (0.209) & 1.003 (0.209)\\
0 & 1 & 0 & 1 & 1 & 0 & 0.759 (0.183) & 0.777 (0.208) & 0.999 (0.208)\\
0 & 1 & 0 & 1 & 1 & 1 & 0.825 (0.187) & 0.779 (0.209) & 1.001 (0.209)\\
0 & 0 & 0 & 1 & 1 & 1 & 1.309 (0.190) & 0.999 (0.218) & 0.999 (0.218)\\
\hline
\end{tabular}
\end{center}
\caption{Means and standard deviations of the estimates of the effect of treatment at visit~2 from an ITT analysis, an estimator of $E[Y_2|Z=1,R=0¨]-E[Y_2|Z=0,R=0]$ and the corrected estimator~(\ref{e:mu2}) for $n_0=n_1=50$, $\sigma_{11}(z)=\sigma_{22}(z)=1$, $\sigma_{12}(z,r)=0.6$ and $c=-0.5$ from 10000 simulations}
\label{t:est}
\end{table}
\par To appreciate the discrepancies between different estimators of a treatment effect of $z$ in the absence of rescue, Table~\ref{t:est} presents the effect estimator from an ITT analysis, an estimator of $E[Y_2|Z=1,R=0¨]-E[Y_2|Z=0,R=0]$ and the estimator corrected according to~(\ref{e:mu2}). The entries are obtained from 10000 simulations per row. Only the corrected estimator is estimating the causal effect $\beta_2$ appropriately.

To make the correction practically useful, the parameters it comprises need to be estimated from the data. Most of them are indeed estimable from observations at visit~1 with the exception of~$\sigma_{12}(z,0)$. With
$$E[Y_1^2(z)|Y_1(z)>c]=\mu_1^2(z)+\sigma_{11}(z)\{\sigma_{11}(z)+[c+\mu_1(z)]\lambda(\eta(z))\}$$
one obtains using similar arguments as in the previous section
\begin{eqnarray*}\label{e:exp12}
\lefteqn{E[Y_1Y_2|Z=z,R=0]}\\
&=&E[Y_1(z)Y_2(z,0)|Y_1(z)>c]\\
&=& E[Y_1(z)E[Y_2(z,0)|Y_1(z)]|Y_1(z)>c]\\
&=& E[Y_1(z)\{\mu_2(z,0)+\gamma_{12}(z,0)(Y_1(z)-\mu_1(z))\}|Y_1(z)>c]\\
&=& \{\mu_2(z,0)-\gamma_{12}(z,0)\mu_1(z)\}E[Y_1(z)|Y_1(z)>c]\\
&& \mbox{}+\gamma_{12}(z,0) E[Y_1^2(z)|Y_1(z)>c]\\
&=& [\mu_2(z,0)-\gamma_{12}(z,0)\mu_1(z)][\mu_1(z)+\sigma_{11}(z)\lambda(\eta(z))]\\
&& \mbox{}+\gamma_{12}(z,0)\{\mu_1^2(z)+\sigma_{11}(z)[\sigma_{11}(z)+(c+\mu_1(z))\lambda(\eta(z))]\}\\
&=& \mu_2(z,0)[\mu_1(z)+\sigma_{11}(z)\lambda(\eta(z))]+\gamma_{12}(z,0)\sigma_{11}[\sigma_{11}(z)+c\lambda(\eta(z))].
\end{eqnarray*}
Substituting $\mu_2(z,0)$ from equation~(\ref{e:mu2}) yields
\begin{eqnarray}
\lefteqn{\sigma_{12}(z,0)=}\nonumber\\
&& \frac{E[Y_1Y_2|Z=z,R=0]-E[Y_2|Z=z,R=0][\mu_1(z)+\sigma_{11}(z)\lambda(\eta(z))]}{1+\lambda(\eta(z))[\eta(z)-\lambda(\eta(z))]}\quad\label{e:sigma12}
\end{eqnarray}
The parameters on the right hand side of (\ref{e:sigma12}) are consistently estimable from the data, hence a consistent estimator of $\sigma_{12}(z,0)$ can be obtained. Data from all patients are used for estimating $\mu_(z)$, $\sigma_{11}$ and $\lambda(\eta(z))$. Only data from patients that did not require rescue medication during the trial are used for $E[Y_1Y_2|Z=z,R=0]$ and $E[Y_2|Z=z,R=0]$.

We run a small simulation study to get an idea of the performance of a plug-in estimator obtained by replacing the parameters on the right-hand side of the preceding equation by their estimates. The results are displayed in Table~\ref{t:est2}.
\begin{table}[t]
\begin{center}
\begin{tabular}{cccccc|ccc}
\hline\hline
$\alpha_1$ & $\beta_1$ & $\alpha_2$ & $\beta_2$ & $\gamma$ & $\delta$ & $\hat\beta_2$ & $\hat\sigma_{12}(0)$ & $\hat\sigma_{12}(1)$\\
\hline
1 & 0 & 0 & 0 & 0 & 0 & 0.001 (0.215) & 0.592 (0.149) & 0.591 (0.151)\\
0 & 1 & 0 & 0 & 0 & 0 & -.006 (0.252) & 0.596 (0.228) & 0.591 (0.149)\\
0 & 0 & 1 & 0 & 0 & 0 & -.003 (0.289) & 0.560 (0.279) & 0.561 (0.273)\\
0 & 0 & 0 & 1 & 0 & 0 & 1.019 (0.297) & 0.597 (0.229) & 0.555 (0.282)\\
0 & 0 & 0 & 0 & 1 & 0 & 0.002 (0.301) & 0.597 (0.229) & 0.596 (0.226)\\
0 & 0 & 0 & 0 & 0 & 1 & -.005 (0.299) & 0.594 (0.228) & 0.597 (0.226)\\
0 & 0 & 0 & 0 & 1 & 1 & 0.002 (0.300) & 0.596 (0.227) & 0.592 (0.229)\\
0 & 1 & 0 & 1 & 0 & 0 & 1.002 (0.257) & 0.596 (0.226) & 0.588 (0.136)\\
0 & 1 & 0 & 1 & 1 & 0 & 0.997 (0.257) & 0.595 (0.227) & 0.584 (0.140)\\
0 & 1 & 0 & 1 & 1 & 1 & 1.001 (0.259) & 0.599 (0.226) & 0.586 (0.138)\\
0 & 0 & 0 & 1 & 1 & 1 & 1.017 (0.292) & 0.597 (0.227) & 0.557 (0.279)\\
\hline
\end{tabular}
\end{center}
\caption{Means and standard deviations of the estimates of the effect of treatment at visit~2 and the correlation of $Y_1(z)$ and $Y_2(z,0)$ for $n_0=n_1=50$, $\sigma_{11}(z)=\sigma_{22}(z)=1$, $\sigma_{12}(z,r)=0.6$ and $c=-0.5$ from 10000 simulations}
\label{t:est2}
\end{table}
As can be seen, the proposed estimate works quite well in the given scenario, however, there seems to be a tendency to underestimate the covariance in particular when the proportion of rescue is increasing. When we re-run the simulation with a higher sample size of 500 in each treatment group this phenomenon practically disappeared. In any case this suggests that the estimate may become inaccurate for small sample sizes.
\section{Discussion}
An estimator of the biological effect of an investigational treatment over control has been proposed for clinical trials where patients are administered rescue medication if the assigned treatment fails. The main assumptions are a bivariate normal model for the potential outcomes at visits 1 and 2, and a deterministic rule for administration of rescue medication based on observations obtained at visit 1. It should be noted that the decision on rescue at visit 1 can be based on an endpoint different from the one on which the efficacy assessment is based on at visit~2. A generalization to more visits should be feasible.

When estimating the biological effect at visit 2, data from all subjects from visit 1 are used in the analysis. However, for visit 2, only observations from subjects without rescue medication are analyzed. The smaller effective sample size results in a somewhat larger variability of the estimate as compared to the ITT estimate which uses the information of the full sample. However, though the latter is using more data, it is not an appropriate biological effect estimator from an accuracy perspective.

As a recommendation for the analysis of clinical trials requiring rescue medication one should conduct an ITT analysis to obtain an estimator of treatment effectiveness, an analysis of the amount and/or timing of rescue medication, and an analysis of the biological effect. If an investigational treatment plus rescue has no or little advantage over control plus rescue, one may wonder what could constitute an additional benefit of the new treatment. Likewise, the efficacy of the investigational treatment is questionable if more rescue is taken there as compared to control. Nevertheless in such a case investigational treatment may still do better than control in patients that would not need rescue under either treatment. To figure that out, an estimate of the biological effect would be helpful. In short there is not one analysis that covers all aspects of the issue. An analysis based on subjects that did not need rescue in there respective groups without any correction is discouraged.

The standard error of the biological estimates can be obtained by re-sampling. More importantly the smaller sample size in principal stratum 00 has to be taken into consideration during the planning phase of a trial. Furthermore, we did not account for missing data in this paper but rather assumed that all data needed to obtain estimates of biological effects are available.
\end{document}